\documentclass{JHEP3} 
\usepackage{graphicx,amssymb,amsfonts,hangcaption}

\makeatletter
\newcounter{subeqncnt}
\def\thesubeqncnt{\alph{subeqncnt}}
\def\subequations{\begingroup%
   \stepcounter{equation}\edef\@tempa{\theequation}%
   \let\c@equation\c@subeqncnt\c@subeqncnt\z@
   \edef\theequation{\@tempa\noexpand\thesubeqncnt}}

\makeatother

\newcommand{\captionfonts}{\small}
\makeatletter 
\long\def\@makecaption#1#2{%
\vskip\abovecaptionskip
\sbox\@tempboxa{{\captionfonts #1: #2}}%
\ifdim \wd\@tempboxa >\hsize
{\captionfonts #1: #2\par}
\else
\hbox to\hsize{\hfil\box\@tempboxa\hfil}%
\fi
\vskip\belowcaptionskip}
\makeatother 

\newcommand{\del}{\partial}

\newcommand{\dd}{{\rm d}}
\newcommand{\ee}{{\rm e}}
\def\ba{\begin{eqnarray}}
\def\ea{\end{eqnarray}}
\def\la{\langle}\def\ra{\rangle}

\title{
 Quark Number Susceptibility with Finite Chemical Potential
in Holographic QCD }

\author{Youngman Kim
        \\
Asia Pacific Center
for Theoretical Physics, Pohang, Gyeongbuk 790-784, Korea \\
Department of Physics, Pohang University of Science and Technology,
Pohang, Gyeongbuk 790-784, Korea
\\  E-mail: \email{ykim@apctp.org}}
\author{Yoshinori Matsuo
 \\
 Harish-Chandra Research Institute,
 Chhatnag Road, Jhusi, Allahabad 211019, India
 \\ E-mail: \email{ymatsuo@hri.res.in}
}

\author{Woojoo Sim, Shingo Takeuchi, Takuya Tsukioka
        \\
Asia Pacific Center
for Theoretical Physics,  Pohang, Gyeongbuk 790-784, Korea
\\  E-mail: \email{space, shingo, tsukioka@apctp.org}}

\preprint{{\tt APCTP-Pre2010-001}\\
{\tt HRI/ST/1002}\\
{\tt arXiv:1001.5343[hep-th]}
}

\abstract{
We study the quark number susceptibility in holographic QCD
with a finite chemical potential
or under an external magnetic field at finite temperature.
We first consider the quark number susceptibility with the chemical
potential.
We observe that approaching $T_c$ from high temperature regime,
$\chi_q/T^2$  develops  a peak as we increase the chemical potential, 
which confirms recent lattice QCD results.
We discuss this behavior in connection with the
existence of the critical end point in the QCD phase diagram.
We also consider the quark number susceptibility under the external
magnetic field. 
We predict that the quark number susceptibility exhibits a
blow-up behavior at low temperature as we raise the value of 
the magnetic field.
We finally spell out some limitations of our study. }


\begin{document}

\section{Introduction}

Understanding the QCD phase diagram is one of the important problems
in modern theoretical physics; see~\cite{QCDphaseDreviews} for some
recent reviews.
An important feature of the QCD phase diagram is the existence
of the critical end point (CEP)
where the first order phase transition, from the hadronic phase to the
quark gluon plasma (QGP), ends.
Based on the fact that at the critical point thermodynamic
observables such as susceptibilities diverge and the order parameter
fluctuates on long wavelengths,
basic ideas to observe CEP are suggested in~\cite{SRS}.

One of the important signals of QGP formation is the fluctuation
of conserved charges such as baryon number or
electric charge~\cite{Mc87, chargeF}.
The quark (or baryon) number susceptibility, which measures the
response of QCD to a change of the quark chemical potential
is one of them~\cite{Mc87, GLTRS}.
There have been many studies to calculate the quark number
susceptibility in various model studies~\cite{Kunihiro:91, CMT, HKRS}
and lattice simulations~\cite{GLTRS88, GPS, GKetal, EKR}.
In \cite{jkls}, the quark number susceptibility at finite temperature
is studied with AdS/QCD models.
The existence of a peak in the quark number susceptibility near
$T_c$ is confirmed by recent lattice QCD calculations based on the
Taylor expansion with respect to the quark (or baryon) chemical
potential ~\cite{Alltonetal, Ejiri}.
This implies the existence of CEP, at which the first order phase
transition terminates in the $(\mu_q, T)$ plane of the QCD phase diagram,
see \cite{CEP} for a review.
Lattice QCD estimates that the location of the CEP would be
$T^E=165-175 ~{\rm MeV}$ and $\mu_B^E=250-400~{\rm MeV}$~\cite{CEP}.
The existence of the CEP  has been also investigated
in various QCD-based model studies~\cite{CEPtheo}.

The AdS/CFT correspondence \cite{Maldacena,gkp,w} is
a powerful tool to investigate strongly coupled gauge theories
including critical phenomena.
Using this correspondence, we can obtain
physical quantities in gauge theories from calculations in gravity side.
Although the correspondence between QCD and gravity theory is not known,
we can obtain much insights by using this correspondence.
In fact, many properties are shared by
various gauge theories, especially in hydrodynamic limit.
Hydrodynamic properties can be read off by introducing
a small perturbations into the bulk fields \cite{ss,PSS}.
This gives small fluctuations to the
fluid in the gauge theory side.
Physical quantities, like transport coefficients,
can be calculated by considering the responses to
these small perturbations; see~\cite{gmsstt, mstty} and references
therein.

In order to study dynamics of quarks (or baryons)
in the gauge theory side, we have to introduce
the $U(1)$ baryon symmetry in the gauge theory side.
According to the Gubser-Klebanov-Polyakov/Witten relation \cite{gkp,w},
a global symmetry in the gauge theory side corresponds to
a gauge symmetry in the gravity side.
To consider the $U(1)$ baryon symmetry,
we have to introduce a $U(1)$ gauge field in the bulk.
This kind of models can be constructed
from the string theory setup.
In D3/D7 setup, for example, D7-branes are considered as
the flavor brane and gauge fields on D7-branes
play the role of the bulk gauge field.
This model has $\mathcal N=2$ supersymmetry, and
hence, the corresponding gauge theory
is $\mathcal N=2$ supersymmetric QCD (SQCD).
Even though the real QCD does not have supersymmetry,
this discrepancy is expected to be solved by
universality mentioned above.

In this paper, we study one of the promising QGP probes by using
the AdS/CFT correspondence.
The primary goal is to calculate the quark number susceptibility
at finite temperature with a finite quark chemical potential.
Our approach has some ups and downs.
Contrary to the lattice QCD considerations,
we can calculate the susceptibility with arbitrary values of
the chemical potential.
However, we are not able to study the quark number susceptibility
in confined phase, which will be discussed at the end of 
the section~\ref{qSUSRNAdSHW}.
Moreover, our study based on AdS/CFT is inherently suffering
from  $1/N_c$ corrections.
Unfortunately, a systematic way of collecting all those corrections
has not been established. With this caution in mind,
we first revisit the quark number susceptibility at finite temperature,
and then we generalize it with the chemical potential.
For this we work in the AdS black hole and Reissner-Nordstr\"om-AdS
backgrounds.

In addition to this, we evaluate the quark number susceptibility
at finite temperature with a constant magnetic field.
The basic motivation is due to the observation that
the constant magnetic field enhances
the dynamical chiral symmetry breaking~\cite{mCat}.
On top of it, recently it is argued that sufficiently large magnetic
fields are created in heavy ion collisions~\cite{KMW},
and so our study may be tested in a terrestrial experiment.
Therefore it is interesting to study the phase diagram of QCD
as a function of the magnetic field, and so the quark number
susceptibility.

\section{Retarded Green functions and quark number susceptibility}

In this section, we briefly summarize the prescription for
the Minkowskian correlator in the AdS/CFT correspondence,
and then define the quark number susceptibility through the correlator.
We here follow the prescription proposed in~\cite{ss}.
We work on the following 5D background,
\begin{equation}
 \dd s^2 = g_{\mu\nu}\dd x^\mu\dd x^\nu + g_{uu}(\dd u)^2,
\end{equation}
where $x^\mu$ and $u$ are the 4D and radial coordinate,
respectively.
We refer the boundary at $u=0$ and the horizon at $u=1$.
Let us consider a solution of an equation of motion
in this 5D background.
Suppose the solution is given by
\begin{equation}
\phi(u,x) =
\!\int\!\frac{\dd^4 k}{(2\pi)^4}\ \mbox{e}^{ikx}f_k(u)\phi^0(k),
\end{equation}
where $f_k(u)$ is normalized such that $f_k(0)=1$ at the boundary.
After putting the equation of motion back into the action,
the on-shell action might be reduced to surface terms
\begin{equation}
S[\phi^0]
=\!\int\!\frac{\dd^4 k}{(2\pi)^4}
\phi^0(-k){\cal G}(k, u)\phi^0(k)
\bigg|_{u=0}^{u=1} .
\label{on_shell_action}
\end{equation}
Here, the function $\mathcal G(k,u)$
can be written in terms of $f_{\pm k}(u)$ and $\partial_u f_{\pm k}(u)$.
Accommodating
Gubser-Klebanov-Polyakov/Witten relation~\cite{gkp,w}
to Minkowski spacetime,
Son and Starinets proposed
the formula to get the retarded Green functions,
\begin{equation}
G^{\rm R}(k)
=
2{\cal G}(k, u)
\bigg|_{u=0},
\label{green_function}
\end{equation}
where the incoming boundary condition at the horizon is imposed.
In this paper, we consider correlators of $U(1)$ currents $J_\mu(x)$,
where $J_\mu (x)$ is the vector current of quark field
or quark number current.
Now we define the precise form of the retarded Green functions
which we discuss later:
\begin{equation}
G_{\mu\ \nu}(k)
=-i\!\int\!\dd^4x \ \ee^{-ikx}\theta(t)
\langle [J_\mu (x), \ J_\nu (0)]\rangle.
\label{cc}
\end{equation}

The quark number susceptibility was proposed as a probe of the
QCD chiral phase transition at zero chemical potential~\cite{Mc87, GLTRS},
\begin{equation}
\chi_q=\frac{\del n_q}{\del\mu_q}.
\end{equation}
Later it has been shown that the quark number susceptibility can
be rewritten in terms of the retarded Green function through
the fluctuation-dissipation theorem~\cite{Kunihiro:91},
\begin{equation}
\chi_q(T, \mu)
=-\lim_{k\to 0} {\mbox{Re}}\Big(G_{t\ t}(\omega=0, k)\Big),
\label{suss}
\end{equation}
where $G_{\mu \ \nu}(\omega, k)$ is the retarded Green function,
which is defined in (\ref{cc}).

\section{Reissner-Nordstr\"om-AdS background}
\label{qSUSRNAdS}

In this section, we briefly review the Reissner-Nordstr\"om-AdS (RN-AdS)
background in 5D spacetime first.
Afterwards, we explain the interpretation of RN-AdS spacetime
in terms of the D3/D7-brane setting.

We consider the Einstein-Maxwell action with negative cosmological
constant:
\begin{equation}
S
=
\frac{1}{2\kappa^2}\!\int\!\dd^5x\sqrt{-g}
\Big(
R-2\Lambda
\Big)
-
\frac{1}{4e^2}
\!\int\!\dd^5x\sqrt{-g}
{\cal F}_{mn}{\cal F}^{mn},
\label{action_bh}
\end{equation}
where we denote the gravitation constant and the cosmological constant
as $\kappa^2=8\pi G_5$ and $\Lambda$, respectively.
The $U(1)$ gauge field strength is given by
${\cal F}_{mn}(x)=\del_m{\cal A}_n(x)-\del_n{\cal A}_m(x)$.
The RN-AdS background is a solution of the equations of motion
derived from this action, and it is given by
\begin{subequations}
\begin{eqnarray}
\dd s^2
&=&
\frac{r^2}{l^2}
\bigg(
-f(r)(\dd t)^2+(\dd \vec{x})^2
\bigg)
+\frac{l^2}{r^2f(r)}(\dd r)^2,
\label{rnads}
\\
{\cal A}_t
&=&
-\frac{Q}{r^2}+\mu,
\label{rnads_1}
\end{eqnarray}
\end{subequations}

\vspace*{-7mm}
\noindent
with
$$
f(r)
=
1-\frac{ml^2}{r^4}+\frac{q^2l^2}{r^6},
\qquad
\Lambda
=
-\frac{6}{l^2}, \qquad
e^2=\frac{2Q^2}{3q^2}\kappa^2.
$$
The parameters $l$, $m$, and $q$ are the AdS radius, mass and
charge, while
$Q$ and $\mu$ are $U(1)$ charge and constant which is interpreted as
the chemical potential.
Since the gauge potential ${\cal A}_t(x)$ must vanish at the horizon,
the charge $Q$ and the chemical potential $\mu$ are related as
\begin{equation}
\mu=\frac{Q}{r_+^2}.\label{ccR}
\end{equation}

The horizons of the RN-AdS black hole are located
at the zero for $f(r)$\footnote{
In order to define the horizon, the charge $q$ must satisfy
a relation $q^4\le 4m^3l^2/27$.
},
\begin{equation}
f(r)
=
1-\frac{ml^2}{r^4}+\frac{q^2l^2}{r^6}
=
\frac{1}{r^6}
\Big(r^2-r_+^2\Big)
\Big(r^2-r_-^2\Big)
\Big(r^2-r_0^2\Big),
\label{metric}
\end{equation}
where the explicit forms of the horizon radii are given by
\begin{subequations}
\begin{eqnarray}
r_+^2
&=&
\left(
\frac{m}{3q^2}
\Bigg(
1+2\cos\bigg(\frac{\theta}{3}+\frac{4}{3}\pi\bigg)
\Bigg)
\right)^{-1},
\label{r+}
\\
r^2_-
&=&
\left(
\frac{m}{3q^2}
\Bigg(
1+2\cos\bigg(\frac{\theta}{3}\bigg)
\Bigg)
\right)^{-1},
\label{r-}
\\
r_0^2
&=&
\left(
\frac{m}{3q^2}
\Bigg(
1+2\cos
\bigg(
\frac{\theta}{3}
+\frac{2}{3}\pi
\bigg)
\Bigg)
\right)^{-1},
\label{r0}
\end{eqnarray}
\end{subequations}

\vspace*{-7mm}
\noindent
with $r^2_++r^2_-=-r^2_0$.
 Here
$$
\theta
=
\arctan
\Bigg(
\frac{3\sqrt{3}q^2\sqrt{\displaystyle 4m^3l^2-27q^4}}{2m^3l^2-27q^4}
\Bigg)\, .
$$
The positions expressed by $r_+$ and $r_-$ correspond to the outer
and inner horizon, respectively.
It is useful to notice that the charge
$q$ can be expressed in terms of $\theta$ and $m$ by
$$
q^4=\frac{4m^3l^2}{27}\sin^2\left(\frac{\theta}{2}\right).
$$
The outer horizon takes a value in
$$
\sqrt{\frac{m}{3}}l
\le r_+^2 \le \sqrt{m}l,
$$
where the upper bound and the lower bound correspond to the case for
$q=0$ and the extremal case, respectively.

The temperature is defined from the conical singularity free
condition around the horizon $r_+$,
\begin{equation}
T
=
\frac{r_+^2f'(r_+)}{4\pi l^2}
=
\frac{r_+}{\pi l^2}
\bigg(
1-\frac{1}{2}\frac{q^2l^2}{ r_+^6}
\bigg)
\equiv
\frac{1}{2\pi b}
\Big(
1-\frac{a}{2}
\Big),
\quad (>0),
\label{temp}
\end{equation}
where
\begin{equation}
a\equiv\frac{q^2l^2}{r_+^6}, \qquad
b\equiv\frac{l^2}{2r_+}.
\end{equation}
In the limit $q\rightarrow 0$,  these parameters go to
$$
a\rightarrow 0,  \qquad
b\rightarrow  \frac{l^{3/2}}{2m^{1/4}},
$$
and the temperature becomes
$$
T\rightarrow T_0
= \frac{m^{1/4}}{\pi l^{3/2}}.
$$
It might be useful to rewrite the parameters
$a$ and $b$ in terms of the temperature and the chemical potential,
\begin{equation}
a = 2 - \frac{4}{1+\sqrt{1+4(\tilde{\mu}/T)^2}}, \qquad
b = \left(\frac{1}{\pi T}\right) \frac{1}{1+\sqrt{1+4(\tilde{\mu}/T)^2}},
\label{a_and_b}
\end{equation}
where we denote
$\tilde{\mu}\equiv\mu\sqrt{8\pi G_5/(3(\pi el)^2)}$. 

For later convenience,
we shall introduce new dimensionless coordinate
$u\equiv r_+^2/r^2$ which is normalized by the outer horizon.
In this coordinate system, the horizon and the boundary are located at
$u=1$ and $u=0$, respectively.
The background metric (\ref{rnads}) can be rewritten as
\begin{equation}
\dd s^2
=\frac{l^2}{4b^2u}\Big(-f(u)(\dd t)^2+(\dd\vec{x})^2\Big)
+\frac{l^2}{4u^2f(u)}(\dd u)^2,
\label{rnads_01}
\end{equation}
with
$$
f(u)=(1-u)(1+u-au^2).
$$

Now let us explain the interpretation of this RN-AdS
spacetime in terms of the brane setting.
We consider $N_c$ D3-branes and treat them
as a gravitational background i.e. AdS$_5\times S^5$.
In order to consider the baryon charge at the boundary theory,
we introduce $N_f$ D7-branes.
The D7-branes are wrapping on $S^3$ of $S^5$, and
we neglect this $S^3$ dependence.
Here we use the probe approximation for the D7-branes,
so that a backreaction from ``the D7-brane tension'' is neglected.
Then the effective action becomes that for 5D gauge theory.
The baryon current corresponds to $U(1)$ gauge field on
this flavor D7-branes.
The 5D effective action of the system might be given by
the action (\ref{action_bh}) with
\begin{equation}
\frac{l^3}{\kappa^2}=\frac{N_c^2}{4\pi^2}, \qquad
\frac{l}{e^2}=\frac{N_cN_f}{4\pi^2}.
\end{equation}
The baryon charge which can be identified by $Q$
and the charge of RN-AdS $q$ are then related
by~\cite{SJS}
\begin{equation}
q^2=\frac{2}{3}\frac{N_f}{N_c}l^2Q^2.
\label{qQ}
\end{equation}
By using the relation (\ref{qQ}),
one might understand that AdS leads RN-AdS through
a backreaction from ``the $U(1)$ baryon charge'' on D7-branes.
However this does not necessarily imply that $N_c$ and $N_f$ are of
the same order of magnitude.
The baryon charge is provided by open strings between D3- and D7-branes.
We can introduce large numbers of these strings even for small $N_f$.
Then, the geometry receives the backreaction from the charge
even though the D7-branes itself is treated as a probe.
We here consider the case in which $N_f$ is finite while the
charge $Q$ is large.
This can be understood in the large $N_c$ context through the
relation (\ref{qQ}).
The finite contribution of the charge $q$ of
RN-AdS could be only considerable if the large baryon
charge $Q(\propto\sqrt{N_c})$ is taken in the large $N_c$.

We will evaluate the quark number susceptibility using the
hard wall model \cite{EKSS, PR} and soft wall model~\cite{KKSS} 
with the RN-AdS background.
Note that the Hawking-Page type transition in both models 
is studied in~\cite{Herzog}.
In the hard wall model to ensure the confinement 
a sharp infrared (IR) cutoff is introduced,
while in the soft wall model a dilaton-like field is coupled 
to the 5D bulk action.
At high temperature, due to the Hawking-Page type transition 
discussed in \cite{Herzog}, AdS black hole background is stable. 
In this case the black hole horizon completely covers up
the IR cutoff of the hard wall model, while in the soft wall 
model an IR scale, which is associated with the dilaton-like field, 
is still visible. 
So the only energy scale in the hard wall model at high 
temperature is the temperature itself.
In the soft wall model, however, we have two energy scales, 
temperature and the IR scale. 
A consequence of this is that the quark number susceptibility 
in the hard wall model turns out to be  temperature-independent, while
that in the soft wall model shows non-trivial temperature 
dependence at high temperature. 
The latter is close to the observations made in lattice QCD and 
also in QCD models.

\subsection{Quark number susceptibility in hard wall model}
\label{qSUSRNAdSHW}

In this subsection, we shall discuss the quark number susceptibility
in the hard wall model on the RN-AdS background.
The 5D action of the gauge field, which is dual to the 4D quark
number current $j_\mu(x)=\bar q(t,\vec x) \gamma_\mu q(t,\vec x)$,
is
\ba
S =-\frac{1}{4g_5^2}\!\int\!\dd^5x\sqrt{-g}F_{mn}F^{mn},
\label{actionHard}
\ea
where $g_5$ is the 5D gauge coupling constant.
In this work we consider two different values of
the gauge coupling constant:
$g_5^2 =12 \pi^2 /N_c$ from the hard wall model~\cite{EKSS, PR}
and $g_5^2=4 \pi^2 l/(N_c N_f)$ from D3/D7.

To obtain the quark number susceptibility,
we use the action (\ref{actionHard}) with
the metric and gauge perturbations around the classical
background (\ref{rnads}) and (\ref{rnads_1}).
Since the Green function which provides the quark number
susceptibility (\ref{suss})
is the current-current correlator $G_{t\ t}(k)$,
we here need to consider the scalar type in the metric perturbation.
We follow the procedure in~\cite{mstty} and refer to the result
\begin{equation}
G_{t\ t}(\omega, k)
=-\frac{lk^2}{4g_5^2(1+a)b^2}
\Big\{
\frac{9a}{k^2-3\omega^2}
+\frac{2(2-a)^2b}{D_{\rm p}(\omega, k)}
\Big\},
\end{equation}
where
$$
D_{\rm p}(\omega, k)
=2 (2+a) b k^2-4 i (1+a) \omega- (2-a)^2 b D_- \omega^2,
$$
with
\begin{eqnarray*}
D_-
&=&
\frac{2}{(2-a)^4 (1+4a)^{3/2}}
\Bigg\{
-27 (2-a) a^2 \sqrt{1+4a}
\nonumber\\
&&
\hspace*{37mm}
+4 (1+4a)^{3/2} (1+a)^3 \log (2-a)
\nonumber
\\
&&
\hspace*{37mm}
+4 (2-2a+41a^2)(1+a)^2 K_1(1)\Bigg\},
\\
K_1(1)
&=&
\frac{1}{2}\log(2-a)
-\log\Big(1-\frac{2a}{1+\sqrt{1+4a}}\Big).
\end{eqnarray*}
By using the formula (\ref{suss}), we obtain
\begin{equation}
\chi_q(T, \mu)
=
\frac{l}{2g_5^2b^2}\left(\frac{2+5a}{2+a}\right),
\label{cQH}
\end{equation}
where $a$ and $b$ are given through the definition (\ref{a_and_b}),
\ba
b^{-1} = \pi T \Bigg(
	1+\sqrt{1+\frac{16 l}{3(N_c g_5)^2}
	\bigg(\frac{\mu}{T}\bigg)^2}
	\Bigg),\,
\qquad
a &=2-4\pi T b.
\ea
In the charge free case $\mu=0$, our result agrees with
that in \cite{jkls}.

Before going on further with the quark number susceptibility,
we briefly discuss the Hawking-Page type transition.
As in~\cite{Herzog}, there exists Hawking-Page type transition 
in the hard wall model and soft wall model.
At low temperature in confined phase, thermal AdS dominates 
the partition function,
while at high temperature in deconfined phase, 
AdS black hole geometry dominates.
Therefore, the quark number susceptibility is described
by the AdS black hole background at high temperature
and by the thermal AdS at low temperature.
The critical temperature for deconfinement is
$\sim 120 ~{\rm MeV}$ in the hard wall model and
$T_c\sim 190 ~{\rm MeV}$ in the soft wall model~\cite{Herzog}.
A similar critical temperature was estimated
as $T_c\sim 210 ~{\rm MeV}$ by using the soft metric model~\cite{av}.
Note that the value of the critical temperature depends on
the number of quark flavors, quark mass and quark number
density~\cite{HPT_RNAdS, HPt_more}.
To obtain the critical temperature in the present case,
we have to do the Hawking-Page type analysis with
charged thermal AdS and RN-AdS backgrounds.
Since the charged thermal AdS background has not been discovered,
we could not precisely fix the critical temperature
for the deconfinement transition.
For simplicity, we assume that the critical temperature is
around the value obtained in~\cite{Herzog}:
$T_c= 100 ~{\rm MeV}$ in the hard wall model and
$T_c= 200 ~{\rm MeV}$ in the soft wall model.

The result in (\ref{cQH}) is shown in Figure~\ref{fig:hw},
where the gauge coupling of D3/D7 has been used. We confirmed that
the result with  the gauge coupling from the hard wall model shows 
a similar behavior.
Below $T_c$ we adopt the quark number susceptibility calculated
in the thermal AdS background~\cite{jkls}.
As well known, the thermal AdS would not exhibit the temperature 
dependence.
Therefore, the quark number susceptibility would be a temperature 
independent constant, and it turns out zero~\cite{jkls}.
For high temperature regime, we use different values of the chemical
potential, $\mu=0.0\sim 0.8$ GeV.
Approaching $T_c$ from high temperature regime,
we observe in Figure~\ref{fig:hw} that
$\chi_q/T^2$  shows  a blow-up behavior as we increase $\mu$,
which may indicate the existence of the CEP in the QCD phase diagram.
Compared to lattice QCD, our study has an advantage
that we do not need to do a Taylor expansion with respect to
the chemical potential.
However, our study has a limitation that we cannot study the
temperature dependence of the quark number susceptibility below
the critical temperature,
which is a generic problem in a model study based on
the AdS/CFT due to large $N_c$ nature.
Also, the RN-AdS is describing a QCD-like or  QGP-like system.
For instance the relation between the chemical potential and its conjugate
charge (or number) density is given by $Q\sim \mu T^2$, see (\ref{ccR}).
In a realistic system like QGP, the relation 
in general takes the following form $Q\sim a \mu^3+ b\mu T^2$, 
where $a$ and $b$ are constants. 
%
\FIGURE{
\centering
\includegraphics[scale=1]{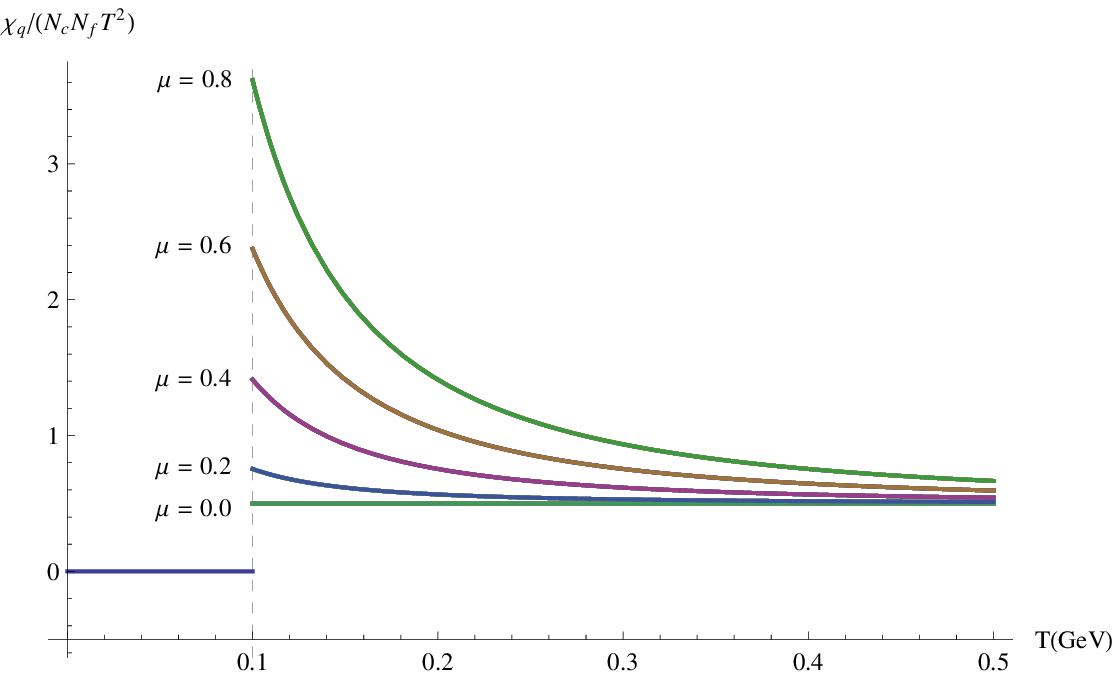}
\caption{$\chi_q/(N_cN_fT^2)$ in the hard wall model for
varying $\mu$(GeV) with $N_c=3$ and $N_f=2$.}
\label{fig:hw}
}

\subsection{Quark number susceptibility in soft wall model}

We shall work in the soft wall model~\cite{KKSS}.
We consider the following action
\begin{equation}
S = -\frac{1}{4g^2_5} \!\int\! \dd^5 x \sqrt{-g}\
\ee^{-\Phi} F_{mn} F^{mn},
\label{action_soft}
\end{equation}
with the dilaton-like field $\Phi(x)$ together with $U(1)$
gauge field $A_m(x)$.
We make use of the RN-AdS geometry (\ref{rnads}) and (\ref{rnads_1})
as the background.

In~\cite{KKSS}, the dilaton like field is given
by $\Phi(x)=l^4\tilde{c}/r^2$.
Rewriting it in terms of $u$, we have
\begin{equation}
\Phi(x)=cu,
\end{equation}
where $c\equiv l^4\tilde{c}/r_+^2$.
We shall work in $A_u(x)=0$ gauge and use the Fourier decomposition
\begin{equation}
A_\mu(t, z, u)=\!\int\!\frac{\dd^4k}{(2\pi)^4}\
\ee^{-i\omega t+ikz}A_\mu(k, u),
\label{fourier}
\end{equation}
where we choose the momenta which are along the $z$-direction.
Equations of motion with respect to $A_t(u)$ and $A_u(u)$
are given as,
\begin{subequations}
\begin{eqnarray}
0&=&
A_t''+\frac{(\ee^{-\Phi})'}{\ee^{-\Phi}}A_t'
-\frac{b^2}{uf}\Big(k^2A_t+k\omega A_z\Big),
\label{eom_t}
\\
0&=&
\omega A_t'+kfA_z',
\label{eom_u}
\end{eqnarray}
\end{subequations}

\vspace*{-6mm}
\noindent
where the prime stands for the derivative with respect to $u$.
An equation of motion for $A_z(u)$ can be derived from
(\ref{eom_t}) and (\ref{eom_u}).
For $A_x(u)$ and $A_y(u)$, one can obtain decoupled second order
ordinary differential equations.
Since we are interested in the time-time component of the retarded
Green function to calculate the quark number susceptibility,
we will not consider $A_x(u)$ and $A_y(u)$ hereafter.

\subsubsection{Solution for $A_t$}

From the equations (\ref{eom_t}) and (\ref{eom_u}),
we obtain an equation of motion
\begin{equation}
0=A_t'''+\frac{(\ee^{-\Phi}uf)'}{\ee^{-\Phi}uf}A_t''
+\Big(
\frac{b^2}{uf^2}(\omega^2-k^2f)-\frac{(\Phi'uf)'}{uf}\Big)A_t'.
\label{eom_t_00}
\end{equation}
Since the equation (\ref{eom_t_00}) is an ordinary second order
differential equation for $A_t'(u)$ with a regular singularity at the
horizon $u=1$,
we first write the solution as $A_t'(u)=(1-u)^\nu F(u)$ where $F(u)$ might
be a regular function at the horizon.
The parameter $\nu$ could be fixed as
\begin{equation}
\nu=-i\frac{\omega}{4\pi T},
\end{equation}
by imposing the incoming wave condition.

Now we  solve the equation of motion in the hydrodynamic
regime i.e.\ small $\omega$ and $k$ compared with the temperature $T$.
In order to do the perturbative analysis
it might be useful to factorize the dilaton factor from $F(u)$,
\begin{equation}
F(u)=\ee^{\Phi(u)}\widetilde{F}(u),
\end{equation}
so that the equation (\ref{eom_t_00}) can be expressed as
a simpler form
\begin{eqnarray}
0
&=&
\Big(\ee^\Phi uf\widetilde{F}'\Big)'
\nonumber
\\
&&
+i\omega\frac{2b}{2-a}\ee^\Phi u(1+u-au^2)\widetilde{F}'
+i\omega\frac{b}{2-a}\Big(\ee^\Phi u(1+u-au^2)\Big)'\widetilde{F}
\nonumber
\\
&&
+\frac{\omega^2b^2}{(2-a)^2}\frac{\ee^\Phi}{1+u-au^2}
\Big((2-a)^2+(1-a)(3-a)u
\nonumber
\\
&&
\hspace*{43mm}
+(1-4a+a^2)u^2
-a(2-a)u^3+a^2u^4\Big)\widetilde{F}
\nonumber
\\
&&
-k^2b^2\ee^\Phi\widetilde{F}.
\label{eom_t_01}
\end{eqnarray}
The function $\widetilde{F}(u)$ is now expanded as
\begin{equation}
\widetilde{F}(u)
=
F_0(u) + \omega F_\omega(u) + k^2 F_{k^2}(u)
+ \mathcal{O}(\omega^2, \omega k^2),
\label{expand_t_01}
\end{equation}
where the functions $F_0(u)$, $F_\omega(u)$ and $F_{k^2}(u)$
are determined by imposing suitable boundary conditions.
The solution can be obtained recursively\footnote{
The derivation of the solutions is given in Appendix \ref{at_in_soft}.
}.
The result is as follows\footnote{
The function $Ei(x)$ is an exponential integral
$$
Ei(x)=-\!\int^\infty_{-x}\!\!\! \dd t\ \frac{\ee^{-t}}{t}.
$$
This can be expanded as follows:
$$
E_i(x)
=
\gamma + \log x + x +{\cal O}(x^2),
$$
where $\gamma$ is the Euler constant.
}:
\begin{subequations}
\begin{eqnarray}
F_0(u)
&=&
C, \qquad (\mbox{const.})
\\
F_\omega (u)
&=&
iCb\ee^c
\Bigg\{
E_i(-cu)+K_1(u)-K_1(0)
\nonumber
\\
&&
\hspace*{14mm}
-\frac{\ee^{-c}}{2-a}
\Big(E_i(c(1-u))-E_i(c)-\log(1-u)\Big)
\Bigg\},
\\
F_{k^2}(u)
&=&
-\frac{Cb^2\ee^c}{c}
\Bigg\{
E_i(-cu)+K_1(u)-K_1(0)
\nonumber
\\
&&
\hspace*{18mm}
-\frac{\ee^{-c}}{2-a}
\Big(E_i(c(1-u))-E_i(c)-\log(1-u)\Big)
\nonumber
\\
&&
\hspace*{18mm}
-\frac{(1+a)\ee^{-c}}{2(2-a)\sqrt{1+4a}}
\Bigg(
\log
\left(
\frac{\displaystyle 1+\frac{1-2au}{\sqrt{1+4a}}}
{\displaystyle 1+\frac{1}{\sqrt{1+4a}}}
\right)
-\log
\left(
\frac{\displaystyle 1-\frac{1-2au}{\sqrt{1+4a}}}
{\displaystyle 1-\frac{1}{\sqrt{1+4a}}}
\right)
\Bigg)
\nonumber
\\
&&
\hspace*{18mm}
-\ee^{-c}\log u
+\frac{(1-a)\ee^{-c}}{2(2-a)}\log(1+u-au^2)
\Bigg\},
\end{eqnarray}
\end{subequations}

\vspace*{-6mm}
\noindent
with
\begin{eqnarray*}
K_1(u)
&=&
\frac{1}{2(2-a)\sqrt{1+4a}}
\Bigg\{
\ee^{-\frac{c}{2a}(1+\sqrt{1+4a})}
\Big((1+a)-(1-a)\sqrt{1+4a}\Big)
\nonumber \\
&&
\hspace*{45mm}\times E_i\Big(\frac{c}{2a}(1+\sqrt{1+4a}-2au)\Big)
\nonumber
\\
&&
\hspace*{34mm}
-\ee^{-\frac{c}{2a}(1-\sqrt{1+4a})}
\Big((1+a)+(1-a)\sqrt{1+4a}\Big)
\nonumber \\
&&
\hspace*{45mm}\times E_i\Big(\frac{c}{2a}(1-\sqrt{1+4a}-2au)\Big)
\Bigg\}.
\end{eqnarray*}

Let us consider the integration constant $C$.
This could be estimated in terms of the boundary values of the fields
$$
\lim_{u\to 0}A_t(u)=A_t^0, \qquad
\lim_{u\to 0}A_z(u)=A_z^0.
$$
Using the equation of motion (\ref{eom_t}), a relation
\begin{equation}
\lim_{u\to 0}\Big(uf(A_t''-\Phi' A_t')\Big)
=b^2\Big(k^2 A_t^0+\omega kA_z^0\Big)
\end{equation}
should  hold.
Therefore we may fix the constant $C$ as
\begin{equation}
C=\frac{b\Big(k^2A_t^0+\omega kA_z^0\Big)}
{\ee^c\Big(\displaystyle i\omega-\frac{b}{c}(1-\ee^{-c})k^2\Big)}.
\end{equation}
One can see the existence of the hydrodynamic pole in the complex
$\omega$-plane.

Near the boundary the obtained solution leads a
relation between the radial derivative of the fields
and its boundary values
\begin{eqnarray}
A_t'(u)
&=&
b^2\Big(k^2A_t^0+\omega k A_z^0\Big)\log u
\nonumber
\\
&&
+\frac{1}{\ee^c\left(\displaystyle i\omega-\frac{b}{c}(1-\ee^{-c})k^2\right)}
\Bigg\{b\Big(k^2A_t^0+\omega k A_z^0\Big)+{\cal O}(\omega k^2,
k^4)\Bigg\} +{\cal O}(u).
\end{eqnarray}
One can also obtain a similar relation for $A_z'(u)$ through
the equation (\ref{eom_u}).

\subsubsection{Retarded Green functions}

We now proceed to evaluate the Minkowski correlators.
An on-shell action can be obtained from (\ref{action_soft}),
\begin{equation}
S_0[A]
=\frac{l}{4g_5^2b^2}
\!\int\!\frac{\dd^4k}{(2\pi)^4}
\ \ee^{-\Phi(u)}
\Bigg(
A_t(-k, u)A_t'(k, u)
-f(u)A_z(-k, u)A_z'(k, u)
\Bigg)\Bigg|_{u=0}^{u=1}.
\end{equation}
By using the relation (\ref{green_function}) and the definition
(\ref{cc}), we can read off the correlators in the hydrodynamic
approximation,
\begin{subequations}
\begin{eqnarray}
G_{t \ t}(\omega, k)
&=&
\frac{l}{2g_5^2b}\frac{k^2}
{\displaystyle \ee^{c}\left(i\omega-Dk^2\right)},
\label{soft_tt}
\\
G_{t \ z}(\omega, k)
&=&
-\frac{l}{2g_5^2b}\frac{\omega k}
{\displaystyle \ee^{c}\left(i\omega-Dk^2\right)},
\\
G_{z \ z}(\omega, k)
&=&
\frac{l}{2g_5^2b}\frac{\omega^2}
{\displaystyle \ee^{c}\left(i\omega-Dk^2\right)},
\label{soft_zz}
\end{eqnarray}
\end{subequations}

\vspace*{-6mm}
\noindent
where we have introduced the following local counter term in the boundary
to remove the logarithmic singularity:
$$
S_{\rm ct}
=\frac{l}{8g_5^2}\log\varepsilon
\!\int\!
\dd^4x\sqrt{-g^{(4)}}F_{\mu\nu}F^{\mu\nu}.
$$
The constant $D$ is the diffusion constant
\begin{equation}
D=\frac{b}{c}(1-\ee^{-c}).
\end{equation}
Using the correlator (\ref{soft_tt}),
we can obtain the quark number
susceptibility $\chi_q$ in terms of the temperature and the chemical
potential through the definition (\ref{suss}),
\begin{equation}
\chi_q (T, \mu)
=\frac{l}{2g_5^2b^2}\left(\frac{c}{\ee^{c}-1}\right).
\label{chiQTmu}
\end{equation}
In the charge free case it reduces to
\ba
\chi_q (T)
=\frac{2\pi^2T^2}{g_5^2}\left(\frac{c}{\ee^{c}-1}\right),
\label{chiQwT}
\ea
which is different from~\cite{jkls}.
We confirmed that
(\ref{chiQwT}) is correct starting from the 5D AdS-Schwarzschild background.
In terms of $\tilde c (=cr_+^2/l^4)$, the susceptibility
(\ref{chiQTmu}) is given by
\ba\label{qns_sw}
\chi_q(T, \mu) = \frac{2l\tilde c}{g_5^2(\ee^{4b^2 \tilde c}-1)}.
\ea
In the soft wall model~\cite{KKSS}, $\tilde c$ is fixed by the rho
meson mass.
In the present case we cannot use hadronic observables such as masses
or couplings to fix the constant since we are working in the black hole
phase, where light mesons are to be melted away.
Here we take another route to fix it.
We compare our $c_2$ defined below, equation (\ref{exp}),
with that from lattice QCD~\cite{Alltonetal} at $T=T_c$
and choose $\tilde c$ to reproduce the lattice result. 
\ba\label{exp}
\chi_q/T^2=\sum_n 2n(2n-1)c_{2n}(\mu/T)^{2(n-1)}.
\ea

The quark number susceptibility with the chemical potential
is shown in Figure~\ref{fig:sw},
where the gauge coupling from D3/D7 has been used.
For high temperature regime, we use different values of the
chemical potential, $\mu=0.0\sim 1.6$ GeV.
Again we find that the quark number susceptibility shows a blow-up
behavior as we lower the temperature to the critical temperature,
thereby indicating the existence of the CEP in the $(\mu, T)$ plane
QCD phase diagram.
%
\FIGURE{
\centering
\includegraphics[scale=1]{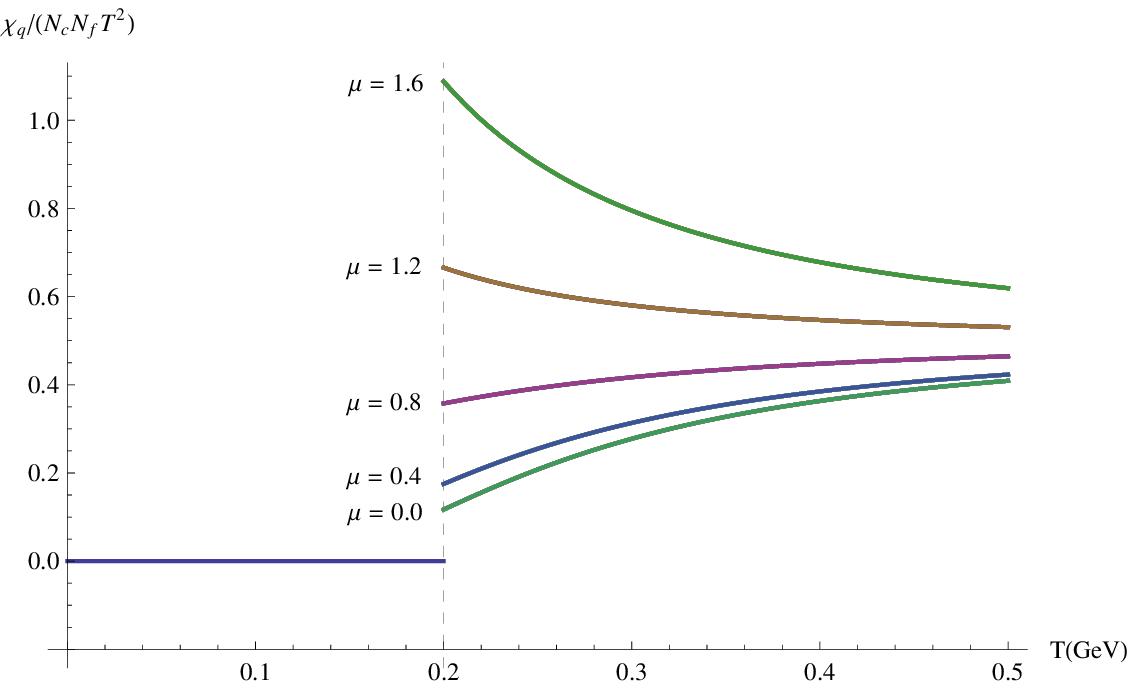}
\caption{$\chi_q/(N_cN_fT^2)$ in the soft wall model for
varying $\mu$(GeV) with $N_c=3$ and $N_f=2$.}
\label{fig:sw}
}

\subsection{Implication in QCD phase diagram}
\label{QCDPd}

Here we discuss our results in the light of QCD phase diagram.
As well known, the nature of the QCD transition does depend 
on the number of quark flavors and the value of the quark mass.
For pure $SU(3)$ gauge theory with no quarks, it is the first order.
In the case of two massless and one massive quarks,
the transition is the second order at zero or small quark chemical 
potentials, 
and it becomes the first order as we increase the chemical potential. 
The point where the second order transition becomes
the first order is called tricritical point.
With physical quark masses of up, down, and strange, 
the second order at zero or low chemical potential
becomes the crossover, and the tricritical point turns into 
the critical end point, see \cite{CEP} for reviews.

Now what can we say about the QCD phase structure based on our study?
Our approach is based on AdS/CFT, and so it inherently probes 
QCD (or QCD-like theory) at large $N_c$.
The transition suggested by our study is manifestly first order 
as shown in the figures, \ref{fig:hw} and \ref{fig:sw}, 
where the quark number susceptibility shows a discontinuous 
jump at the transition point.
This discontinuity is there since we use thermal AdS at low 
temperature and RN-AdS at high temperature.
This first order nature of the transition would be the end of
story as long as we don't consider $1/N_c$ corrections.
Though we may not be able to assemble all the leading $1/N_c$ 
corrections in a consistent way, we discuss some speculation 
on what could come out with those corrections. 
For zero chemical potential, as shown in Figs. \ref{fig:hw} 
and \ref{fig:sw}, we have two lines at low  and high temperature
regimes. 
There is a hope that with  $1/N_c$ corrections those two straight 
lines are connected with no discontinuity at zero or small chemical 
potentials since the discontinuous gap is not much big.
If this happens, the transition could be the second order or the crossover.
When we increase the chemical potential, this possibility does not 
seem plausible since the quark number susceptibility in high 
temperature regime blows up near the transition point 
and also the transition point will move to the left, i.e.\  
the transition temperature decreases with the chemical 
potential~\cite{HPT_RNAdS, HPt_more}. 
This means that the quark number susceptibility at low 
temperature, which is zero with no $1/N_c$ corrections, 
should increase very rapidly to realize smooth change at the 
transition temperature. 
With $1/N_c$ corrections, this will not be possible.
Therefore at high chemical potential, the first order nature of 
the transition will persist regardless of the presence of $1/N_c$ 
corrections, while for small chemical potentials it may change to 
the second order or the crossover due to the $1/N_c$ corrections. 
However, we emphasize here again that this is just a speculation.

In short, the transition from our study is the first order, and it may, 
however, become the second order or the crossover with $1/N_c$ 
corrections.

\section{Quark number susceptibility under magnetic field}

In this section, we study the quark number susceptibility
with an external magnetic field turned on.
The basic motivation is due to the observation that the
constant magnetic field enhances the dynamical chiral symmetry breaking
 $\la\bar qq \ra \sim |eB|$
and generates the dynamical quark mass $m_q^{\rm dyn}=f(|eB|)$~\cite{mCat}.
Using an effective low energy QCD model, linear sigma model with quarks,
the authors of \cite{FM08} showed that with increasing 
magnetic field the QCD transition changes from the crossover to 
the first order.
This implies that there exists the CEP in $(B, T)$ plane as we 
raise the magnetic field.
Therefore, we would expect that the external magnetic field affects
the behavior of the quark number susceptibility at finite temperature.
For instance, the presence of the CEP means a peak in the value of 
quark number susceptibility as we increases the magnetic field. 
So the main motivation of our study with the magnetic field is to observe
if the peak appears with the magnetic field.
In addition, a recent study \cite{KMW} shows that
sufficiently large magnetic fields are likely created in relativistic
heavy ion collisions, and so our study may be tested in experiments.

To calculate the quark number susceptibility at finite temperature, 
we consider here the non-extremal ${\rm AdS}_5\times S^5$,
\ba
\dd s^2=\frac{l^2 (\pi T)^2}{u}
	\Big(-f(u) (\dd t)^2+(\dd\vec{x})^2 \Big)
	+\frac{l^2}{4u^2 f(u)}(\dd u)^2 + l^2 \dd\Omega_5^2,
\ea
where $f(u)=1-u^2$ and $T$ is the temperature.
The gauge field comes from the probe D7-brane
whose action reads
\ba
S_{\rm D7}= -N_f T_7\!\int\!\dd^8x\ \ee^{-\phi}
	\sqrt{-\det({\cal G}_{MN}+2\pi\alpha'{\cal F}_{MN})},
\ea
where $T_7=1/((2\pi)^7l_s^8)$ and $\ee^{\phi}=g_s$.
${\cal G}_{MN}(x)$ is the induced metric which we consider
as the trivial one.
Here the external magnetic field $B$ enters the action as~\cite{magF}
\ba
{\cal F}_{MN}=F_{MN}^{(0)}+F_{MN},\qquad F_{xy}^{(0)}=B.
\ea
Then, wrapping the D7-brane on $S^3$ and taking~\footnote{
One may be tempted to use the hard wall or
soft wall model for simplicity.
In this case, however, the magnetic field does not affect
the equations of motion for gauge fields.
}
\ba
E_{MN}={\cal G}_{MN}+F_{MN}^{(0)},
\ea
we get the 5D action~\footnote{
Since $(E_5^{-1})^{mn}$ is not symmetric,
there exist additional terms in the action,
which are the powers of $(E_5^{-1})^{mn}F_{nm}$.
However, our choice of 4-momentum, $k^\mu=(\omega,0,0,k)$,
makes those terms vanish.
}
\ba
S_{\rm 5D} =-\frac{1}{4g_5^2}
\!\int\!\dd^4x\ \dd u\sqrt{-\det(E_5)}\
(E_5^{-1})^{ml}(E_5^{-1})^{kn}F_{mn}F_{lk},
\ea
where $g_5=\sqrt{4\pi^2 l/(N_c N_f)}$ and
we have defined an inverse of 5D part of $E_{MN}$
i.e.\ $(E_5^{-1})^{ml}E_{5ln}=E_{5nl}(E_5^{-1})^{lm}=\delta^m_n$.
As we did in the previous section,
we take the gauge $A_u(x)=0$ and the same Fourier decomposition
as (\ref{fourier}).
The resulting equations of motion with respect to $A_t(u)$ and $A_u(u)$
lead 
\begin{subequations}
\begin{eqnarray}
0&=&
X_{tz}^{-1}\big(X_{tu} A_t'\big)'-(k^2 A_t+k\omega A_z),
\label{eom_b_t}
\\
0&=&
\omega A_t'+ kf A'_z,
\label{eom_b_u}
\end{eqnarray}
\end{subequations}

\vspace*{-7mm}
\noindent
where we have defined
\ba\label{xx}
X_{tz}=\sqrt{-\det(E_5)}(E_5^{-1})^{tt}(E_5^{-1})^{zz},
\qquad
X_{tu}=\sqrt{-\det(E_5)}(E_5^{-1})^{tt}(E_5^{-1})^{uu}.
\ea
Differentiating the equation (\ref{eom_b_t}) with respect to $u$ and
using the equation (\ref{eom_b_u}),
we obtain
\ba
0=\Big(X_{tz}^{-1}\big(X_{tu} A_t'\big)'\Big)'
-\big(k^2-\omega^2 f^{-1}\big) A_t'.
\ea

Now using the hydrodynamic expansion
$$
A_t'(u)=(1-u)^\nu\Big(F_0(u) + \omega F_\omega(u) + k^2 F_{k^2}(u)
+ {\cal O}(\omega^2, \omega k^2)\Big),
$$
where $\nu=-i\omega/(4\pi T)$ as the incoming wave condition,
we get the equations of motion for $F_0(u)$ and $F_{k^2}(u)$, respectively,
\begin{subequations}
\begin{eqnarray}
0&=&(X_{tz}^{-1} (X_{tu} F_0)')',
\label{eom_f0}
\\
0&=&(X_{tz}^{-1} (X_{tu} F_{k^2})')'-F_0.
\label{eom_g1}
\end{eqnarray}
\end{subequations}

\vspace*{-7mm}
\noindent
In (\ref{eom_f0}),
$X_{tu} F_0(u)$ should be a constant $(\equiv C_0)$ to avoid the singularity
at $u=1$ due to $X_{tz}^{-1}(u) \to 0$ as $u \to 1$.
As a result, we obtain
\ba
F_0 = C_0 X_{tu}^{-1}.
\ea
Using this solution, the equation (\ref{eom_g1}) is recasted as
\ba\label{1st_int}
X_{tz}^{-1} (X_{tu} F_{k^2})'= C_0 S(u)
\qquad {\mbox{with}} \qquad
S(u)=\!\int_1^u\! \dd u' X_{tu}^{-1}(u').
\ea
Then, we insert the above solutions into (\ref{eom_b_t}) to obtain
\ba\label{for_zeroth}
k^2 A_t+k\omega A_z
&=& X_{tz}^{-1}\big(X_{tu} A_t'\big)'
\nonumber
\\
&=&
X_{tz}^{-1}\big(X_{tu} k^2 F_{k^2}\big)' + {\cal O}(\omega)
\nonumber
\\
&=&
k^2 C_0 S(u) + {\cal O}(\omega).
\ea
Since we will take $\omega=0$, this equation determines $A_t(u)$.
Note that $A_t(u)$ obtained from this procedure is
the zeroth order term of the series solution,
since we are substituting $A_t''(u)$ evaluated to ${\cal O}(\omega,k^2)$.
In fact, we should integrate the equation (\ref{1st_int}) once more
to get the solution of ${\cal O}(k^2)$, which cannot be done analytically.
However, to get the susceptibility,
only the zeroth order solution is needed
since $k \to 0$ limit of Green function with $\omega=0$
implies the contribution of the zeroth order only.
Thus, from (\ref{for_zeroth}), the zeroth order solution is now
\ba
A_t(u) = C_0 S(u) = A_t^0\frac{S(u)}{S(0)},
\ea
where $A_\mu^0\equiv A_\mu(u)|_{u=0}$.

As a result, we get the retarded Green function and
the quark number susceptibility
\ba\label{x}
\chi_q
=-\lim_{k\to 0}{\mbox{Re}}\Big(G_{t\ t}(\omega=0, k)\Big)
= \frac{1}{g_5^2}[S(0)]^{-1}.
\ea
Then, using the explicit form of $X_{tu}(u)$ from (\ref{xx}),
we obtain
\ba \label{qnsb}
\chi_q(T,B) &=& \frac{1}{g_5^2}
    \bigg[\int_{1}^{0} \frac{\dd u}{X_{tu}}\bigg]^{-1}
\nonumber \\
&=& \frac{1}{g_5^2}
	\bigg[\int_{1}^{0}
		\frac{-l\dd u}{2\sqrt{(2\pi\alpha' B)^2 u^2+(\pi l T)^4}}
	\bigg]^{-1}
\nonumber \\
&=& \frac{4\pi\alpha' B}{g_5^2 l}
	\bigg[{\rm arcsinh}
		\bigg(\frac{2\pi\alpha' B}{(\pi l T)^2}\bigg)
	\bigg]^{-1}.
\ea
When $B=0$, we observe that $\chi_q/T^2\sim T^0$ (constant),
which agrees with the result of the hard wall model at $\mu=0$.
This should be so since the hard wall model action used is
nothing but the leading term of Dirac-Born-Infeld action
for the D7 probe brane after perturbative expansions of the  action
in terms of $\alpha^\prime$.

The plot of $\chi_q/T^2$ for varying 
${\widetilde B}$($\equiv 2\pi\alpha'B$) is given in
Figure~\ref{fig:bbg}, where we take $\tilde B=(0,1,2,3,4)$ from 
bottom to top. 
We find that the quark number susceptibility increases rapidly 
with increasing $\tilde B$ as we lower $T$ from high temperature regime. 
Note that since D3/D7 model does not exhibit the confined phase, 
we plot our quark number susceptibility in entire finite temperature 
regime.
This observation itself is very new, though the blow-up behavior 
at low temperature could be expected by the study of \cite{FM08}. 
By studying the magnetic field dependence of a modified potential, 
the authors of \cite{FM08}
showed that with increasing magnetic field the QCD transition 
changes from the crossover to the first order, which implies
the existence of the CEP, and so the diverging behavior of the 
quark number susceptibility.
In this sense, our study support indirectly the result obtained 
in \cite{FM08}.
But, our study made in this section would not address the QCD 
transition itself due to the absence of the confined phase in D3/D7 model.
To improve this defect and to see if the blow-up behavior is universal 
regardless of the gravity background,
we may consider D4/D6 or D4/D8 model, which is relegated to a future 
study including the effect of a finite quark mass.
Note that the finite quark mass seems soften the peak in the quark 
number susceptibility~\cite{sqSUS}.
\FIGURE{
\centering
\includegraphics[scale=0.7]{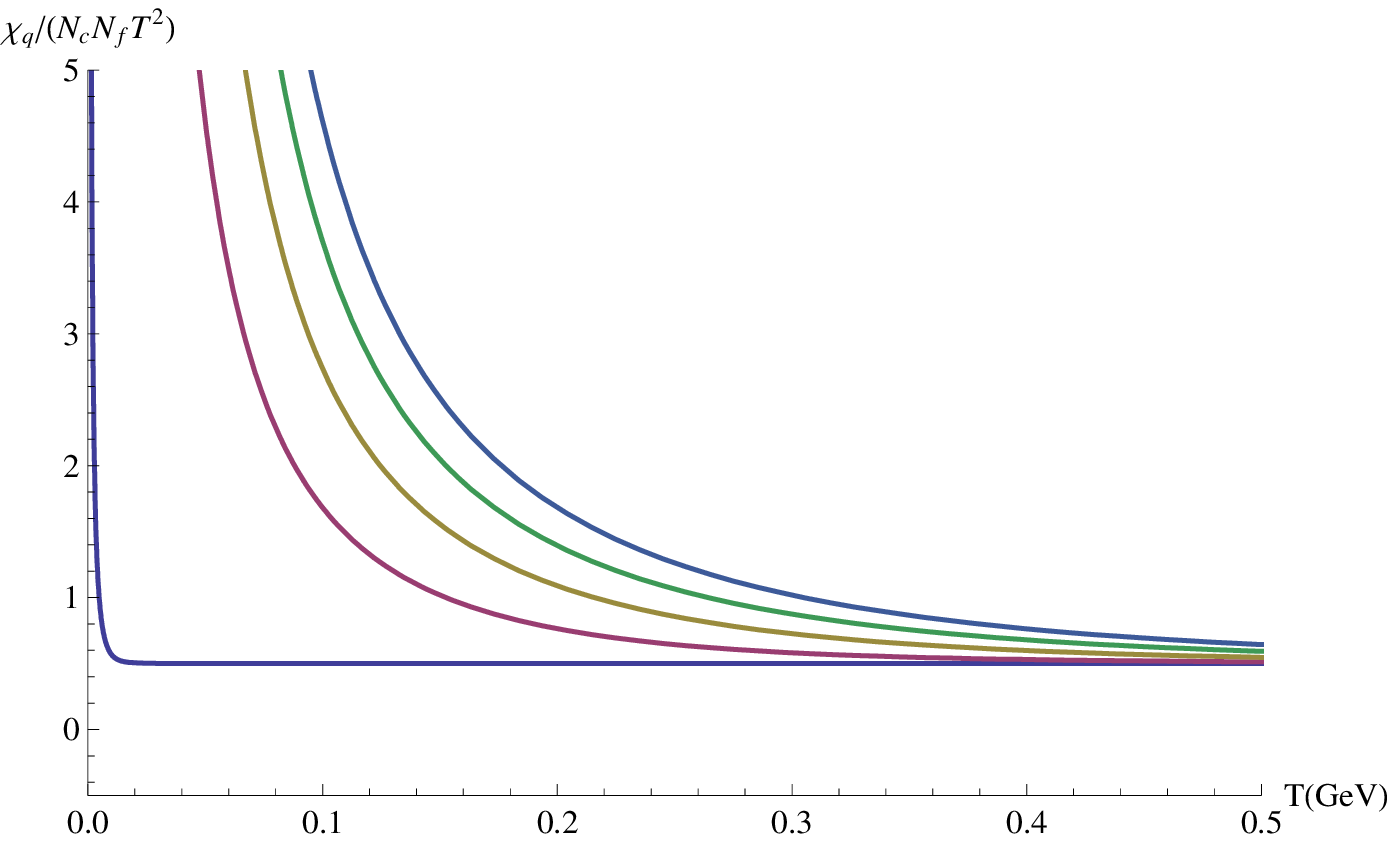}
\caption{$\chi_q/(N_cN_fT^2)$ for varying ${\widetilde B}=2\pi\alpha'B$
with $N_c=3$ and $N_f=2$. Here $\tilde B=(0,1,2,3,4)$ from bottom to top.}
\label{fig:bbg}
}

Before closing this section,
we compare $\chi_q/T^2$ in D3/D7 and D4/D8~\cite{s-s} at finite 
temperature with no magnetic field.
We prepare a basic setup to calculate the quark number susceptibility
with an external magnetic field in D4/D8 in Appendix \ref{AppD4D8}.
In \cite{jk}, from the point of view of the chiral symmetry breaking,
an external magnetic field was considered in D4/D8.
Although we do not provide the result in Appendix B,
we can see the behavior resulting from
D4/D8 system by look through the $B=0$ case.
In that case, Chern-Simons terms do not
contribute and we can easily get the result
\ba
\chi_q(T)
\sim \bigg[\int_{1}^{0}
		\frac{-\dd u}{\sqrt{T^6 u^{-1}}}
	\bigg]^{-1}
\sim T^3.
\ea
This result is, however, different from what we obtained in D3/D7 system.
The different $T$ dependence of $\chi_q$ come from
the different $T$ dependence of the horizons and
from the different exponents of $u$ in the integrands.
The dimensionful parameter $l$ compensates the different powers of $T$.
On the baryonic density and susceptibilities in D4/D8 model,
we refer to \cite{KKim}.

\section{Summary}

We studied the quark number susceptibility with the finite quark
chemical potential or under the external magnetic field at high
temperature.
We first considered the hard wall model in the RN-AdS background and
observed that as we lower the temperature starting from a high
temperature, the quark number susceptibility shows a peak with
increasing $\mu$. When $\mu=0$, however, the quark number susceptibility is
independent of the temperature both at low and high temperatures.
To improve this at least in high temperature regime, 
we move to the soft wall model and found that at high temperature 
the quark number susceptibility increases with the temperature 
for $\mu=0$, and observed a similar blow-up behavior as we lowered 
the temperature with increasing chemical potential.
This peak may imply the existence of the
CEP in QCD phase diagram on $(\mu, T)$ plane.
We discussed implication of our results to the QCD phase diagram 
in \ref{QCDPd}, where we concluded that the transition from our study 
is the first order, and it may, however, become the second order or 
the crossover with $1/N_c$ corrections.

We then calculated the quark number susceptibility under
the external magnetic field.
As we raise the magnitude of the magnetic field,
we observed  a similar rising-up behavior as we lowered the temperature 
with increasing magnetic field. 
It will be interesting if this behavior in $(B, T)$ plane is to be
confirmed or disconfirmed by lattice QCD or any other studies.

Finally, we compared our D3/D7 model study with D4/D8 at high
temperature to find 
\begin{eqnarray*}
\chi_q\sim T^2, &\qquad {\rm D3/D7},
\\
\chi_q\sim T^3, &\qquad {\rm D4/D8}.
\end{eqnarray*}
We observed that the $T$ dependence of $\chi_q$ is sensitive to 
the background geometry. 
In QGP, $\chi_q/T^2$ saturates to a constant value, 
ideal gas limit, at sufficiently high temperature.
It is interesting to see that D3/D7 model shows this feature.
This might imply that the blow-up behavior in $(\mu, T)$ or $(B, T)$ 
plane may vary with the gravity background.
Therefore it will be interesting to see if the blow-up behavior 
is universal.
On top of this, finite quark mass effect on the quark number 
susceptibility is also interesting to investigate.
These issues will be addressed in a future study.

Before closing the paper, we remark some limitations of our study.
In our study,
we can calculate the susceptibility with arbitrary values
of the chemical potential.
However, we are not able to study the temperature dependence 
of the quark number susceptibility
in confined phase due to the Hawking-Page type transition. 
Therefore, we evaluate the quark number susceptibility only
in high temperature regime, deconfined phase.
In addition, our study based on AdS/CFT is
inherently suffering from $1/N_c$ corrections.
Unfortunately, a systematic way of collecting all those corrections
has not been established.

\acknowledgments

YK and TT acknowledge the Max Planck Society (MPG),
the Korea Ministry of Education,
Science, Technology (MEST), Gyeongsangbuk-Do and Pohang City for the support
of the Independent Junior Research Group at the Asia Pacific Center for
Theoretical Physics (APCTP).

\appendix

\section {Perturbative solution of $A_t$ in
the soft wall model\label{at_in_soft}}

Plugging the expansion (\ref{expand_t_01}) into the equation
(\ref{eom_t_01}), one can read off the one for $F_0(u)$,
\begin{equation}
0=\Big(\ee^\Phi ufF_0'\Big)'.
\end{equation}
A general solution is given by
\begin{equation}
F_0(u)
=
C_0
+D_0
\Big\{
E_i(-cu)
-\frac{\ee^{-c}}{2-a}E_i(c(1-u))
+K_1(u)
\Big\},
\end{equation}
where $C_0$ and $D_0$ are integration constants.
Imposing the regular condition at the horizon,
the solution should be
\begin{equation}
F_0(u)=C_0\equiv C. \qquad (\mbox{const.})
\end{equation}
The constant $C$ could be estimated later.

By using this solution, one can get an equation for $F_\omega(u)$,
\begin{equation}
0=\Big(\ee^\Phi ufF_\omega'\Big)'
+i\omega\frac{b}{2-a}\Big(\ee^\Phi u(1+u-au^2)\Big)'C.
\end{equation}
A general solution is
\begin{eqnarray}
F_\omega(u)
&=&
C_1
+D_1
\Big\{
E_i(-cu)
+K_1(u)
\Big\}
\nonumber
\\
&&
-\frac{\ee^{-c}}{2-a}
\Big\{D_1 E_i(c(1-u))-iCb\ee^c\log(1-u)\Big\}.
\end{eqnarray}
Removing the singularity at the horizon,
the integration constant $D_1$ should be
$$
D_1=iCb\ee^{c}.
$$
In order to fix the another constant $C_1$,
it might be convenient to impose a condition at the boundary,
$$
\left[
F_\omega(u)-E_i(-cu)\lim_{u\to 0}\left(\frac{F_\omega(u)}{E_i(-cu)}\right)
\right]_{u=0}\!\!\!\!=0,
$$
so that finite terms at the boundary could be removed.
The final form of the solution is
\begin{eqnarray}
F_\omega(u)
&=&
iCb\ee^c
\Bigg\{
E_i(-cu)+K_1(u)-K_1(0)
\nonumber
\\
&&
\hspace*{14mm}
-\frac{\ee^{-c}}{2-a}
\Big(E_i(c(1-u))-E_i(c)-\log(1-u)\Big)
\Bigg\}.
\end{eqnarray}

A differential equation for $F_{k^2}(u)$ is
\begin{equation}
0=
\Big(\ee^{\Phi}ufF_{k^2}'\Big)'-Cb^2\ee^{\Phi}.
\end{equation}
A general solution can be obtained as
\begin{eqnarray}
F_{k^2}(u)
&=&
C_2+D_2\Big\{E_i(-cu)+K_1(u)\Big\}
\nonumber
\\
&&
-\frac{\ee^{-c}}{2-a}
\Big\{D_2E_i(c(1-u))+\frac{Cb^2\ee^c}{c}\log(1-u)\Big\}
\nonumber
\\
&&
+\frac{Cb^2}{2(2-a)c}
\Bigg\{
\frac{1+a}{\sqrt{1+4a}}
\log\left(
\frac{\displaystyle 1+\frac{1-2au}{\sqrt{1+4a}}}
{\displaystyle 1-\frac{1-2au}{\sqrt{1+4a}}}
\right)
\nonumber
\\
&&
\hspace*{23mm}
+2(2-a)\log u -(1-a)\log(1+u-au^2)
\Bigg\},
\end{eqnarray}
and the constant $D_2$ can be fixed as
$$
D_2=-\frac{Cb^2\ee^c}{c}.
$$
The remaining constant $C_2$ might be determined to eliminate finite
contributions at the boundary.
The solution then becomes
\begin{eqnarray}
F_{k^2}(u)
&=&
-\frac{Cb^2\ee^c}{c}
\Bigg\{
E_i(-cu)+K_1(u)-K_1(0)
\nonumber
\\
&&
\hspace*{18mm}
-\frac{\ee^{-c}}{2-a}
\Big(E_i(c(1-u))-E_i(c)-\log(1-u)\Big)
\nonumber
\\
&&
\hspace*{18mm}
-\frac{(1+a)\ee^{-c}}{2(2-a)\sqrt{1+4a}}
\Bigg(
\log
\left(
\frac{\displaystyle 1+\frac{1-2au}{\sqrt{1+4a}}}
{\displaystyle 1+\frac{1}{\sqrt{1+4a}}}
\right)
-\log
\left(
\frac{\displaystyle 1-\frac{1-2au}{\sqrt{1+4a}}}
{\displaystyle 1-\frac{1}{\sqrt{1+4a}}}
\right)
\Bigg)
\nonumber
\\
&&
\hspace*{18mm}
-\ee^{-c}\log u
+\frac{(1-a)\ee^{-c}}{2(2-a)}\log(1+u-au^2)
\Bigg\}.
\end{eqnarray}
\section{D4/D8 brane system with the external magnetic field \label{AppD4D8}}

In this section we shall discuss an effect of a constant magnetic
field to the quark number susceptibility.
We here consider the D4/D8 brane system with the constant magnetic filed.

We first introduce the bulk background geometry of $N_c$ D4-branes
in type IIA superstring theory with the compactification on a
circle.
There exist two different phases i.e.\ confined and deconfined phases.
We here take the deconfined phase.
The background is then given by
\begin{subequations}
\begin{eqnarray}
\dd s^2
&=&
\left(\frac{U}{R}\right)^{3/2}
\!\!\Big(-f(U)(\dd t)^2+(\dd\vec{x})^2+(\dd x_4)^2\Big)
+\left(\frac{R}{U} \right)^{3/2}
\!\!\left(\frac{(\dd U)^2}{f(U)}+U^2\dd\Omega_4^2
\right), \qquad \
\\
\ee^\phi
&=&
g_s\left(\frac{U}{R}\right)^{3/4},
\end{eqnarray}
\end{subequations}

\vspace*{-7mm}
\noindent
where the $\phi(x)$ is the dilaton field.
$\dd\Omega_4^2$ is the metric of the four sphere and
$R$ is the curvature radius of the background which is expressed
in terms of the string coupling $g_s$ and the string
length $l_s=\sqrt{\alpha'}$,
$$
R^3=\pi g_s N_cl_s^3.
$$
The function $f(U)$ is given by
\begin{equation}
f(U)=1-\left(\frac{U_T}{U}\right)^3,
\end{equation}
and the temperature can be read off as
\begin{equation}
T=\frac{3}{4\pi}\frac{U_T^{1/2}}{R^{3/2}}.
\end{equation}

Following Sakai and Sugimoto~\cite{s-s},
we introduce the probe D8-brane which
sits in the transverse direction to the compactified one $x_4$.
In the deconfined phase where
the $x_4$-$U$ subspace forms a cylinder,
the D8-brane might be a straight line
which simply follows the geodesic from $U=U_T$ to infinity,
\begin{equation}
\dd s^2_{\rm D8}
=
\left(\frac{U}{R}\right)^{3/2}
\!\!\Big(-f(U)(\dd t)^2+(\dd\vec{x})^2\Big)
+\left(\frac{R}{U} \right)^{3/2}
\!\!\left(\frac{(\dd U)^2}{f(U)}+U^2\dd\Omega_4^2
\right).
\label{d8-metric}
\end{equation}

The action for the D8-brane consists of the sum of the
DBI and the Chern-Simons actions.
The DBI action is given by
\begin{equation}
S_{\rm DBI}
=-T_8\!\int\!\dd^9x\ \ee^{-\phi}\sqrt{-\det({\cal G}_{MN}
+2\pi\alpha'{\cal F}_{MN})},
\label{dbi}
\end{equation}
where $T_8=1/((2\pi)^8l_s^9)$ is the D8-brane tension and
${\cal G}_{MN}(x)$ is the induced metric (\ref{d8-metric}).
We put the constant magnetic field $B$ as the background
of the $U(1)$ gauge field in the D8-brane and consider small
fluctuations,
\begin{equation}
{\cal F}_{MN}=F_{MN}^{(0)}+F_{MN} \quad {\mbox{with}} \quad
F_{xy}^{(0)}=B.
\end{equation}
It might be useful to collect the background fields as
\begin{equation}
E_{MN}={\cal G}_{MN}+2\pi\alpha'F_{MN}^{(0)}.
\end{equation}
Integrating over the four-sphere, we can then obtain
the following action for the fluctuations
from the DBI action
(\ref{dbi}),
\begin{equation}
S_{\rm 5D}
=
-\frac{N_cR}{96\pi^3\alpha'}
\!\int\!
\dd^4x\hspace*{0.8mm}
\dd U\sqrt{-\det(E_5)}\hspace*{0.8mm}\left(\frac{U}{R}\right)^{1/4}
(E_5^{-1})^{ml}(E_5^{-1})^{kn}F_{mn}F_{lk},
\end{equation}
where the indices $m$ and $n$ run through $t, 1, 2, 3, U$ and
an inverse of 5D part of $E_{MN}$ has been defined
i.e.\ $(E_5^{-1})^{ml}E_{5ln}
=E_{5nl}(E_5^{-1})^{lm}=\delta^m_n$.
We set the four sphere components of the gauge fields to be zero.
The 5D Chern-Simons action arises after
an integration of the RR four form over the four sphere on the
D8-brane,
\begin{equation}
S_{\rm CS}
=-i\frac{N_c}{48\pi^2}\!\int\!{\cal A}\wedge {\cal F}\wedge {\cal F}.
\end{equation}

As we did in the main part of the paper,
we work on the gauge $A_U(x)=0$ and use the same Fourier
decomposition as (\ref{fourier}).
Equations of motion for $A_t(U)$ and $A_z(U)$ are then given by
\begin{subequations}
\begin{eqnarray}
0
&=&
\alpha
\Bigg\{
\Bigg(\left(\frac{U}{R}\right)g(U)A_t'\Bigg)'
-\left(\frac{U}{R}\right)^{-2}\!\!\frac{g(U)}{f(U)}
\Big(k^2A_t+\omega kA_z\Big)
\Bigg\}
-i\beta BA_z',
\label{eom_ss_cs_t}
\\
0
&=&
\alpha
\Bigg\{
\Bigg(\left(\frac{U}{R}\right)g(U)f(U)A_z'\Bigg)'
+\left(\frac{U}{R}\right)^{-2}\!\!\frac{g(U)}{f(U)}
\Big(\omega^2A_z+\omega kA_t\Big)
\Bigg\}
-i\beta BA_t', \quad
\label{eom_ss_cs_z}
\\
0
&=&
\alpha\left(\frac{U}{R}\right)g(U)
\Big(\omega A_t'+kf(U)A_z'\Big)
-i\beta B\Big(\omega A_z+kA_t\Big),
\label{eom_ss_cs_u}
\end{eqnarray}
\end{subequations}

\vspace*{-6mm}
\noindent
where
\begin{equation}
g(U)=\sqrt{\left(\frac{U}{R}\right)^3+(2\pi\alpha'B)^2}.
\end{equation}
The constants $\alpha$ and $\beta$ are defined by
$$
\alpha=\frac{N_cR}{24\pi^3\alpha'},
\qquad
\beta=\frac{N_c}{8\pi^2}.
$$
The equations (\ref{eom_ss_cs_t}) and (\ref{eom_ss_cs_u}) imply
(\ref{eom_ss_cs_z}).

In order to solve the set of equations,
it is standard to introduce the master variable,
\begin{equation}
Z(U)
=\omega A_z(U) + kA_t(U).
\end{equation}
The the master equation becomes the following form:
\begin{eqnarray}
0
&=&
Z''
+\Bigg(
\frac{1}{U}
+\frac{g'(U)}{g(U)}+\frac{\omega^2f'(U)}{f(U)(\omega^2-k^2f(U))}
\Bigg)Z'
+\left(\frac{U}{R}\right)^{-3}\frac{\omega^2-k^2f(U)}{f^2(U)}Z
\nonumber
\\
&&
-i\frac{\beta B}{\alpha}\left(\frac{U}{R}\right)^{-1}\!\!\!
\frac{\omega kf'(U)}{f(U)g(U)(\omega^2-k^2f(U))}Z
\nonumber
\\
&&
+\frac{\beta^2B^2}{\alpha^2}\left(\frac{U}{R}\right)^{-2}\!\!\!
\frac{Z}{f(U)g^2(U)}.
\label{master_00}
\end{eqnarray}
It might be convenient to introduce a dimensionless radial coordinate
$u\equiv U_T/U$.
The locations of the horizon and the boundary correspond to
$u=1$ and $u=0$, respectively.
By using this coordinate,
the master equation (\ref{master_00}) becomes
\begin{eqnarray}
0
&=&
Z''
+\Bigg(\frac{1}{u}+\frac{g'(u)}{g(u)}
+\frac{\omega^2f'(u)}{f(u)(\omega^2-k^2 f(u))}
\Bigg)Z'
+\left(\frac{3}{4\pi T}\right)^2\frac{\omega^2-k^2f(u)}{uf^2(u)}Z
\nonumber
\\
&&
+i3\pi\alpha'B\frac{\omega kf'(u)}{uf(u)g(u)(\omega^2-k^2f(u))}Z
\nonumber
\\
&&
+(3\pi\alpha'B)^2\frac{Z}{u^2f(u)g^2(u)},
\label{master_01}
\end{eqnarray}
with
$$
f(u)=1-u^3, \qquad
g(u)=\sqrt{\left(\frac{4\pi
RT}{3}\right)^6\frac{1}{u^3}+(2\pi\alpha'B)^2},
$$
where the prime now implies the derivative with respect to $u$.
We can impose the incoming wave condition at the horizon,
\begin{equation}
Z(u)=(1-u)^{-i\frac{\omega}{4\pi T}}F(u),
\end{equation}
where the function $F(u)$ should be regular at the horizon.
The master equation then becomes that for the function $F(u)$,
\begin{eqnarray}
0
&=&
F''
+\Bigg(\frac{1}{u}+\frac{g'(u)}{g(u)}
+\frac{\omega^2f'(u)}{f(u)(\omega^2-k^2 f(u))}
+i\frac{\omega}{2\pi T}\frac{1}{1-u}
\Bigg)F'
\nonumber
\\
&&
+\Bigg\{
i\frac{\omega}{4\pi T}\frac{1}{1-u}
\Big(\frac{1}{1-u}+\frac{1}{u}
+\frac{g'(u)}{g(u)}+\frac{\omega^2f'(u)}{f(u)(\omega^2-k^2f(u))}
\Big)
\nonumber
\\
&&
\hspace*{6mm}
+\left(\frac{1}{4\pi T}\right)^2
\Big(
-\frac{\omega^2}{(1-u)^2}+\frac{9(\omega^2-k^2f(u))}{uf^2(u)}
\Big)
\Bigg\}F
\nonumber
\\
&&
+i3\pi\alpha'B\frac{\omega kf'(u)}{uf(u)g(u)(\omega^2-k^2f(u))}F
+(3\pi\alpha' B)^2\frac{F}{u^2f(u)g^2(u)}.
\end{eqnarray}
Multiplying the factor $(\omega^2-k^2f(u))$ to the equation above,
we could apply the hydrodynamics approximation.
The function $F(u)$ can be expanded as
\begin{equation}
F(u)=F_0(u)+\omega F_\omega(u)+k F_k(u) +{\cal O}(\omega^2, k^2,
\omega k).
\end{equation}
The order ${\cal O}(\omega k)$ in the expansion of the master
equation,
we can fix the function $F_0(u)$ as
\begin{equation}
F_0(u)=0,
\end{equation}
Equations for $F_k(u)$ and $F_\omega(u)$ can be read off from the
${\cal O}(\omega^2 k)$ and ${\cal O}(\omega k^2)$, respectively,
\begin{eqnarray}
0
&=&
\Big(uf(u)g(u)F'_k(u)\Big)'+i3\pi\alpha'Bf'(u)F_\omega(u)
+(3\pi\alpha'B)^2\frac{F_k(u)}{ug(u)},
\\
0
&=&
\Big(ug(u)F'_\omega(u)\Big)'+i3\pi\alpha'B\left(\frac{1}{f(u)}\right)'F_k(u)
+(3\pi\alpha'B)^2\frac{F_\omega(u)}{uf(u)g(u)}.
\end{eqnarray}
%


\end{document}